\documentclass[aps,reprint,amsmath,amssymb,groupedaddress]{revtex4-1}
\usepackage{graphicx}

\begin{document}

%Title of paper
\title{Entanglement-assisted codeword stabilized quantum codes}

\author{Jeonghwan Shin}
\email{jhsh@korea.ac.kr}
\affiliation{School of Electrical Engineering, Korea University, Seoul, Korea}
\author{Jun Heo}
\email{junheo@korea.ac.kr}
\affiliation{School of Electrical Engineering, Korea University, Seoul, Korea}
\author{Todd A. Brun}
\email{tbrun@usc.edu}
\affiliation{Communication Sciences Institute, University of Southern California,
Los Angeles, CA 90089, USA}

\date{\today}

\begin{abstract}
Entangled qubit can increase the capacity of quantum error correcting codes based on stabilizer codes.  In addition, by using entanglement quantum stabilizer codes can be construct from classical linear codes that do not satisfy the dual-containing constraint.  We show that it is possible to construct both additive and non-additive quantum codes using the codeword stabilized quantum code framework.  Nonadditive codes may offer improved performance over the more common stabilizer codes.  Like other entanglement-assisted codes, the encoding procedure acts only the qubits on Alice's side, and only these qubits are assumed to pass through the channel.  However, errors the codeword stabilized quantum code framework gives rise to effective $Z$ errors on Bob side.  We use this scheme to construct new entanglement-assisted non-additive quantum codes, in particular, ((5,16,2;1)) and ((7,4,5;4)) codes.
\end{abstract}

% insert suggested PACS numbers in braces on next line
% insert suggested keywords - APS authors don't need to do this
%\maketitle must follow title, authors, abstract, \pacs, and \keywords
\maketitle

%%%%%%%%%%%%%%%%%%%%%%%%%%%%%%%%%%%%%%%%%
\section{Introduction}

Quantum computation has attracted great interest because efficient algorithms have been found to solve a variety of classical problems, such as factoring, that are believed to be hard for classical computers.  Moreover, as  the processor size in classical computers continues to scale down, the quantum nature of the components of a classical computers will begin to be important.  Performing reliable classical computations on machines built of quantum components is an important problem; the possibility of exploiting quantum effects to achieve remarkable new performance is an even greater incentive to understand these systems.

In quantum computation, it is important to preserve coherence of quantum information.  For this purpose, quantum information must be protected by quantum error-correcting codes (QECCs) from unwanted interactions and quantum noise.  While there are many classical error correction schemes which perform close to the classical channel capacity, it is hard to apply classical error correction schemes directly to QEC because of various properties of quantum system---such as no-cloning, continuous error models, and measurement-disturbance tradeoffs---which do not exist in classical systems.  Despite these differences between classical error correction and QEC, it is still possible to develop quantum error-correcting codes based on classical error-correcting structure. 
Stabilizer codes (developed in \cite{CRSS98,GottesmanThesis} among others),  are analogues of classical additive codes.  This type of code is specified by a stabilizer group, which is an Abelian subgroup of the Pauli group on $n$ qubits.  The code space of a stabilizer code is fixed by this stabilizer group.  That is, it is a joint eigenspace with eigenvalue 1.  Stabilizer codes can be constructed from classical linear codes that satisfy a particular dual-containing constraint.

Recently, a more general framework, codeword stabilized quantum (CWS) codes, was introduced in \cite{Cross:2009jo} which includes both additive and non-additive quantum codes.  CWS codes in standard form are defined by a graph \cite{VandenNest04} and a classical binary code.  An important aspect of the CWS framework is the fact that any Pauli error is equivalent in its effects to an error consisting only of $Z$ operators.  This means that any Pauli error can be treated as a classical binary error.  Using a set of these induced errors as the desired correctable error set, a quantum error-correcting code can be constructed from a corresponding classical binary code, albeit one with a nonstandard error set.

The set of codes that can be expressed in this way includes the stabilizer codes, but many others as well.  These additional CWS codes are non-additive.  Non-additive codes (in principle) can encode a logical state of higher dimension than a stabilizer code with the same length in physical qubits, while protecting it from same number of errors.  This promises potential gains in performance for quantum error correction.  (Note, however, that none of the non-additive codes discovered so far have a minimum distance greater than three.)

Another fairly recent development in the study of QECCs are entanglement-assisted quantum error-correcting codes (EAQECCs) \cite{Bennett96b,Bowen02}. A theory of entanglement-assisted stabilizer codes was developed in \cite{Brun20102006}, describing codes that use entangled bits (ebits) shared between the sender and receiver.  The use of shared entanglement has two significant benefits.  First, shared entanglement allows a code to correct a larger number of errors, which may allow the sender to either send more qubits for a given number of correctable errors, or correct more errors for the same rate of transmission.  Second, allowing the use of shared entanglement permits one to overcome the dual-containing constraint of stabilizer codes.  It was shown in \cite{Brun20102006} that it is possible to construct a QECC from any classical linear binary or quaternary code, whether or not it is dual-containing.  Codes that are dual-containing correspond to standard QECCs (that use no entanglement); codes that are not dual-containing correspond to EAQECCs.

Our aim in this paper is to increase capacity of quantum error-correcting codes by applying ebits to CWS quantum codes.  It will be assumed that Bob's halves of the shared ebits do not suffer from errors, because they do not pass through the channel.  However, the encoding operation must only act on Alice's side.  Based on the CWS framework in standard form, all Pauli errors can be represented by errors consisting only of $Z$ operators.  In our entanglement-assisted codes, this equivalence will give rise to $Z$ error on Bob's qubits as well, and as a result, word operators corresponding to these errors act on both Alice's and Bob's qubits.  The encoding operation, however, must still be applied only to Alice's qubits.  Therefore, we will show that the word operators are equivalent to operators that only act on Alice's qubits.

This paper is organized as follows.  In section 2, we give a brief overview of the Pauli group and review the construction of stabilizer codes, entanglement-assisted quantum codes and codeword stabilized quantum codes.  Section 3 gives a detailed description of our framework for constructing entanglement-assisted quantum codes based on the CWS framework, and presents some new codes constructed by this framework.  Finally, in section 4, we conclude.

%%%%%%%%%%%%%%%%%%%%%%%%%%%%%%%%%%%%%%%%%
\section{Background and notation}

The elements of the Pauli group $\mathcal{G}_n$ are all $n$-fold tensor products of $2\times 2$ Pauli matrices, and are denoted by
\begin{eqnarray}
\mathcal{G}_n\equiv i^m\{I,X,Y,Z\}^{\otimes n}\phantom{1}\textrm{for }m\in\{0,1,2,3\} ,
\end{eqnarray}
where $I$ is the $2\times2$ identity and $X$, $Y$ and $Z$ are the Pauli matrices:
\begin{eqnarray}
I &=& \left[\begin{array}{cc}1 & 0\\0 & 1\end{array}\right] , \phantom{1}\phantom{1}
X = \left[\begin{array}{cc}0 & 1\\1 & 0\end{array}\right] , \nonumber\\ 
Y &=& \left[\begin{array}{cc}0 & -i\\i & 0\end{array}\right] , \phantom{1}
Z=\left[\begin{array}{cc}1 & 0\\0 & -1\end{array}\right] . \nonumber
\end{eqnarray}
The Pauli matrices are Hermitian unitary matrices having eigenvalues equal to $1$ or $-1$.
Any two elements of $\mathcal{G}_n$ either commute or anticommute with each other:
\begin{eqnarray}
\textrm{either}\phantom{1} [A,B]&=&0,\phantom{1} \textrm{if}\phantom{1} AB=BA, \nonumber\\
\textrm{or}\phantom{1} \{A,B\}&=&0,\phantom{1} \textrm{if}\phantom{1} AB=-BA, \nonumber
\end{eqnarray}
for all $A,B\in \mathcal{G}_n$. 
Up to the overall phase ($1,i,-1,-1$), any element $g$ of $\mathcal{G}_n$ can be represented by a binary vector $(\mathbf{v}|\mathbf{u})$ of length $2n$ as follows:
\begin{eqnarray}
g &=& Z^{\mathbf{v}}X^{\mathbf{u}}, \phantom{1} \nonumber\\
&=& Z^{v_1}X^{u_1} \otimes Z^{v_2}X^{u_2} \otimes \cdots \otimes Z^{v_n}X^{u_n} ,
\end{eqnarray}
where $\mathbf{v} = v_1 v_2 \cdots v_n$ and $\mathbf{u} = u_1 u_2 \cdots u_n$ are binary strings of length $n$.

%======================================================
\subsection{Stabilizer codes}

Stabilizer codes are a well-known family of additive quantum error-correcting codes. 
An $[[n,k,d]]$ stabilizer code encodes $k$ logical qubits into $n$ physical qubits.  It is a $2^k$-dimensional subspace $\mathcal{C}$ (the codespace) of the Hilbert space $\mathcal{H}_n\equiv(\mathbb{C}^2)^{\otimes n} \equiv \mathbb{C}^{2^n}$.  The codespace $\mathcal{C}$ is specified as the simultaneous $+1$ eigenspace of a set of $n$ commuting stabilizer generators $g_i$ that generate a stabilizer group $\mathcal{S}=\langle \{g_i\} \rangle$.  The stabilizer group is an Abelian subgroup of the $n$-qubit Pauli group $\mathcal{G}_n$.

The encoding procedure of an $[[n,k,d]]$ stabilizer code is described as follows. 
Consider the initial $n$-qubit state with $m=n-k$ ancilla qubits in the state $|0\rangle^{\otimes m}$
\begin{equation}
\label{eq:stabilizer_initial_state}
|\psi'\rangle=\underbrace{|00\cdots 0\rangle}_{m}|\phi\rangle,
\end{equation}
where $|\phi\rangle$ represent an arbitrary $k$-qubit state. 
The $m$ stabilizer generators of the stabilizer group $\mathcal{S}'$ for this (rather trivial) code are
\begin{eqnarray*}
Z_1 &=& ZII\cdots I \\
Z_2 &=& IZI\cdots I \\
&\vdots& \\
Z_m &=& \underbrace{I\cdots IZ}_m\underbrace{I\cdots I}_k . 
\end{eqnarray*}
We also identify $2k$ logical operators that act on the state $|\psi'\rangle$:
\begin{eqnarray*}
\overline{Z}_i' &=& Z_{m+i} \\
\overline{X}_i' &=& X_{m+i} ,
\end{eqnarray*}
for $i=1,\cdots,k$.
The logical operators commute with the stabilizer generators, and either commute or anti-commute with other logical operators as follows:
\[
[\overline{Z}_i',\overline{Z}_j'] = 0 = [\overline{X}_i',\overline{X}_j'] ,
\]
and
\begin{eqnarray*}
[\overline{Z}_i',\overline{X}_j'] &=& 0\phantom{1}\textrm{for } i\ne j , \\
\{\overline{Z}_i',\overline{X}_j'\} &=& 0\phantom{1}\textrm{for } i=j .
\end{eqnarray*}

We treat this initial state as a stabilizer code in order to understand how codes transform under unitary encoding operations.  For stabilizer codes, encoding is done by unitary operators drawn from the Clifford group, that preserve the Pauli group:  if $g\in\mathcal{G}_n$, then $UgU^\dagger\in\mathcal{G}_n$.  After an encoding operation with unitary operator $U_E$, the operators
\begin{equation}
g_i = U_E Z_i U_E^\dagger
\end{equation}
for $i=1,\cdots,m$ become the stabilizer generators of the stabilizer group $\mathcal{S}$ for the new code space $\mathcal{C}(\mathcal{S})$.  The encoded state $|\psi\rangle=U_E|\psi'\rangle$ is stabilized by the operators in $\mathcal{S}$, and the logical operators on $|\psi\rangle$ are
\begin{eqnarray}
\overline{Z}_j &=& U_EZ_{m+j} U_E^\dagger , \nonumber\\
\overline{X}_j &=& U_EX_{m+j} U_E^\dagger ,
\end{eqnarray}
for $j=1,\cdots k$.  These operators satisfy the same commutation relations as the logical operators of the initial state.  Note that we generally only consider the logical operators to be defined up to being multiplied by some element of the stabilizer; thus, each logical operator is represented by an equivalence class of operators in the normalizer of $\mathcal{S}$.  The normalizer $\mathcal{N}(\mathcal{S})$ of $\mathcal{S}$ is generated by the logical operators and stabilizer generators:
\begin{equation}
N(\mathcal{S})=\langle g_1,\cdots,g_m,\overline{Z}_1,\cdots,\overline{Z}_k,\overline{X}_1,\cdots,\overline{X}_k\rangle.\nonumber
\end{equation}
The minimum distance $d$ of a stabilizer code is defined as the minimum weight of all operators in $N(\mathcal{S})-\mathcal{S}$.  We can think of these as the set of nontrivial logical operators; the lowest weight element of this set is the lowest weight error that cannot be detected by that code.

%======================================================
\subsection{Entanglement-assisted quantum error-correcting codes}

Entanglement-assistance \cite{Bennett96b,Bowen02,Brun20102006} is an elegant method that can increase the capacity of QECCs. Using shared ebits between the sender and receiver, it is possible to increase the minimum distance or the code rate of QECC.  Brun, Devetak and Hsieh \cite{Brun20102006} also showed that by including shared entanglement in the stabilizer formalism, stabilizer codes can be constructed from classical error-correcting codes without satisfying the dual-containing restriction.

Let us briefly review the theory of entanglement-assisted stabilizer codes.  Suppose that there are $c$ pairs of maximally entangled states shared by Alice and Bob. It is assumed that the halves of the $c$ ebits on Bob's side do not suffer from any error, since they do not pass through the channel.  An $[[n,k,d;c]]$ EAQECC encodes $k$ logical qubits into $n$ physical qubits using $c$ ebits.  We can think of constructing an EAQECC in the following way.  In Eq.~(\ref{eq:stabilizer_initial_state}), the initial state has $m=n-k$ ancilla qubits in the state $|0\rangle$.  We can replace $c$ of the ancillas with $c$ halves of maximally entangled pairs in the state $|\Phi_+\rangle=\frac{1}{\sqrt{2}}(|00\rangle+|11\rangle)$.  This makes the initial state
\begin{equation}
\label{eq:ea_initial}
|\psi'\rangle_{EA}=|\Phi\rangle^{\otimes c}|0\rangle^{\otimes (m-c)}|\phi\rangle.\nonumber
\end{equation}
This initial state is fixed by stabilizer group $\mathcal{S}'$ generated by stabilizer generators 
\begin{equation}
\label{eq:EA_stabilizer}
\left\{\begin{array}{cc}
Z_i^A |  Z_i^B, & \textrm{for}\phantom{1}i=1,\cdots,c \\
X_j^A | X_j^B, & \textrm{ for}\phantom{1}j=1,\cdots,c, \\
Z_i^A | I^B, & \textrm{for}\phantom{1} i=c+1,\cdots,m
\end{array} \right.
\end{equation}
where the operators on the left and right of the `$|$' are applied to the qubits on Alice's and Bob's side, respectively.  The logical operators on $|\psi'\rangle_{EA}$ are
\begin{eqnarray}
\label{eq:EA_loggicalZ} &Z_{m+1}^A| I^B&,\cdots,Z_n^A| I^B , \\
\label{eq:EA_loggicalX} &X_{m+1}^A| I^B&,\cdots,X_n^A| I^B ,
\end{eqnarray}
so they have support only on Alice's side.  For convenience, the superscripts A and B will be omitted throughout the rest of this paper if there are no confusion.  An encoding operation has the form $U_E=U^A| I^B$, applying an encoding operation $U^A$ on Alice's side while no operation is applied on Bob's qubit.

For the code space $\mathcal{C}(\mathcal{S})$ encoded by $U_E$, the encoded state is
\[
|\psi\rangle_{EA}=U_E|\psi'\rangle_{EA} .
\]
The stabilizer generators of  $\mathcal{S}$ for $|\psi\rangle_{EA}$ are
\begin{eqnarray}
g_i &=& U^AZ_i^A(U^A)^\dagger| I^B , \nonumber\\
g_j &=& U^AZ_j^A(U^A)^\dagger| Z_j^B , \nonumber\\
h_j &=& U^AX_j^A(U^A)^\dagger| X_j^B ,
\end{eqnarray}
for $i=c+1,\cdots,m$ and $j=1,\cdots,c$, and the logical operators on $|\psi\rangle_{EA}$ are
\begin{eqnarray}
\overline{Z}_i &=& U_E Z_{m+i}U_E^\dagger , \nonumber\\
\overline{X}_i &=& U_E X_{m+i}U_E^\dagger ,
\end{eqnarray}
for $i=1,\cdots k$.

The stabilizer group can also be expressed in terms of two subgroups, the symplectic subgroup $\mathcal{S}_S$ and the isotropic subgroup $\mathcal{S}_I$, with $\mathcal{S} = \langle\mathcal{S}_I,\mathcal{S}_S\rangle$.  These groups are given by
\begin{equation}
\mathcal{S}_I = \langle \{g_i\} \rangle , \ \ \ \ 
\mathcal{S}_S = \langle \{ g_j ,h_j\} \rangle ,
\end{equation}
for $i=c+1,\cdots,m$ and $j=1,\cdots,c$.  We then can express the normalizer $N(\mathcal{S})$ of $\mathcal{S}$ as follows:
\begin{equation}
N(\mathcal{S}) = \mathcal{S}_I \times \langle \overline{Z}_1,\cdots,\overline{Z}_k,\overline{X}_1,\cdots,\overline{X}_k \rangle .
\end{equation}
The minimum distance $d$ of $C(\mathcal{S})$ is the minimum weight of the operators in $N(\mathcal{S}) - \mathcal{S}_I$, just as for standard stabilizer codes.

%======================================================
\subsection{Codeword stabilized quantum codes}

Codeword stabilized (CWS) codes \cite{Cross:2009jo} are a broad class that includes both additive and non-additive quantum codes.  All CWS codes can be represented in a standard form; in this form, they are specified by a graph $G$ and a classical binary code.  We think of the vertices of the graph $G$ as corresponding to the $n$ qubits of the code, and $G$ has an adjacency matrix $A$.  Based on these, we specify a unique base state and a set of word operators.  The unique state is a single stabilizer state, stabilized by a maximal Abelian subgroup of $\mathcal{G}_n$. In standard form, this stabilizer is generated by a set of Pauli operators:
\begin{equation}
g_i=X_iZ^{\mathbf{r}_i} ,
\end{equation}
where $\mathbf{r}_i$ is the $i$th row vector of the adjacency matrix $A$.  We see that for a CWS code in standard form, the base state is therefore a graph state \cite{Schlingemann01}.

The code space of a CWS code is spanned by a set of basis vectors which result from applying the word operators to the base state. Therefore, the dimension of the code corresponds to the number of word operators.  Unlike stabilizer codes, this dimension need not be a power of 2.  Therefore we introduce a different notation for quantum codes:  a quantum code that encodes a $K$-dimensional codespace into $n$ physical qubits with minimum distance $d$ is an $((n,K,d))$ code.  An $[[n,k,d]]$ stabilizer is therefore an $((n,2^k,d))$ code.  The word operators are Pauli operators that anticommute with one or more of the stabilizer generators for the base state.  They therefore map the base state into an orthogonal state.  The span of all these states is the code space.  (Obviously, the word operators must be chosen so that the different basis states are also orthogonal to each other.)  These basis states are also eigenstates of the stabilizer generators, but with some of the eigenvalues differing from $+1$.

An important feature pointed out in \cite{Cross:2009jo} is that any set of correctable errors acting on a codeword of a CWS code in standard form can be represented by another error consisting only of $Z$ operators.  This set of effective errors gives rise to mapping between the set of quantum errors and a set of classical binary errors (generally acting on multiple bits).  The mapping between a Pauli error $E=\pm Z^{\mathbf{v}}X^{\mathbf{u}}$ and a classical binary error is defined by
\begin{eqnarray}
\label{eq:CWS_map}
Cl_{G}(E=\pm Z^{\mathbf{v}}X^{\mathbf{u}})= \mathbf{v}\oplus\bigoplus_{l=1}^nu_l\mathbf{r}_l,
\end{eqnarray}
where $\mathbf{r}_l$ is the $l$th row of the adjacency matrix for ${G}$, and $u_l$ is the $l$th bit of the vector $\mathbf{u}$.  Using this definition, theorem 3 of \cite{Cross:2009jo} may be given that a CWS code in standard form, characterized by a graph $G$ and a classical code $\mathcal{C}_b$, detects errors from a set $\mathcal{E}$ if and only if $\mathcal{C}_b$ detects errors from a set $Cl_{G}(\mathcal{E})$, and for each $E\in\mathcal{E}$,
\begin{eqnarray}
\textrm{either}\phantom{1}Cl_{G}(E) &\neq& 0 , \\
\textrm{or, for each}\phantom{1}i,\phantom{1}Z^{\mathbf{c}_i}E&=&EZ^{\mathbf{c}_i} ,
\end{eqnarray}
where the $\{Z^{\mathbf{c}_i}\}$ are the word operators from $\mathcal{C}_b$, $\mathcal{W} = \{Z^\mathbf{c}\}_{\mathbf{c}\in\mathcal{C}_b}$.  (So we see that the word operators are derived from the codewords of the binary code.)

%%%%%%%%%%%%%%%%%%%%%%%%%%%%%%%%%%%%%%%%%
\section{Non-additive quantum codes with entanglement}

\subsection{Encoding}

In this section, we introduce a framework to construct non-additive quantum codes using $c$ ebits, based on CWS codes in standard form.  Such codes are entanglement-assisted CWS (EACWS) codes.  To include entanglement, we make the unique initial base state $|S'\rangle$ of EA-CWS code an entangled state.  This state is the simultaneous $+1$ eigenspace of a maximal Abelian subgroup $\mathcal{S}'$ of $\mathcal{G}_{n+c}$ such as
\begin{eqnarray}
S|S'\rangle_{EA}=|S'\rangle_{EA}\textrm{\quad for }S\in\mathcal{S}'.
\end{eqnarray}
For simplicity, we take the initial state $|S'\rangle_{EA}$ (before encoding) to consist of $n-k$ qubits in the state $|0\rangle$, which begin on Alice's side, and $c$ maximally-entangled pairs of qubits shared between Alice and Bob:
\begin{eqnarray}
|S'\rangle_{EA}&=&|0\rangle^{\otimes (n-c)}|\Phi_+\rangle^{\otimes c} \nonumber\\
&\equiv&|0\cdots 0\rangle|e_0\cdots e_0\rangle.
\label{initial_state}
\end{eqnarray}
where $|e_0\rangle = |\Phi_+\rangle$. It is assumed that the halves of the $c$ ebits on Bob's side do not suffer from any errors, since they do not pass through the channel. The stabilizer $\mathcal{S}'$ for $|S'\rangle_{EA}$ has a set of generators
\begin{eqnarray}
Z_i|I,\phantom{1}& &\textrm{for }i=1,\cdots,n-c\\
\left.\begin{array}{c}Z_i|Z_j,\\X_i|X_j,\end{array}\right.
& &\textrm{for }i=n-c+1,\cdots,n
\label{initial_generators}
\end{eqnarray}
where $j=i-(n-c)$ and the operators on the left and right of `$|$' act on Alice's and Bob's qubits, respectively.

The basis states of the code are produced by applying word operators to the initial state.  If the code is $K$-dimensional, there must be $K$ word operators, and these must act only on Alice's qubits.  We start by defining word operators $w'_l$ for $l=1,\ldots,K$ that act on the initial state:
\begin{eqnarray}
\label{eq:ECWS_word}
w'_l&=&X^{\mathbf{x}_l}\otimes Z^{\mathbf{v}_l}X^{\mathbf{u}_l}\\
&=&X^{\mathbf{x}}\otimes Z^{\mathbf{v}}X^{\mathbf{u}}|I^{\otimes c}.
\end{eqnarray}
where $\mathbf{x}_l$, $\mathbf{v}_l$ and $\mathbf{u}_l$ are binary vectors.  In the above equation, the identity acting on Bob's qubits means that word operators are only applied on Alice's qubits. 
The number of word operators equals the dimension of the code space.  To encode $K$ logical states, $K$ word operators is required.  The $X^{\mathbf{x}_l}$ operators acts on the first $(n-c)$ qubits in the state $|0\rangle$, while the $Z^{\mathbf{v}_l}X^{\mathbf{u}_l}$ operators act on Alice's halves of the $c$ ebits.  The resulting state is
\begin{equation}
w_l'|S'\rangle_{EA} \equiv |w_l'\rangle =
|\mathbf{x}\rangle\otimes Z^{\mathbf{v_e}}X^{\mathbf{u_e}}|e_0\cdots e_0\rangle .
\end{equation}
The maximum dimension of this code space is $2^{(n+c)}$ (though this obviously would allow for no protection against errors).  Generally, the dimension $K$ is significantly smaller.

%======================================================
\begin{figure}[t]%[h]
  \begin{center}
    \scalebox{0.8}[0.8]{\includegraphics{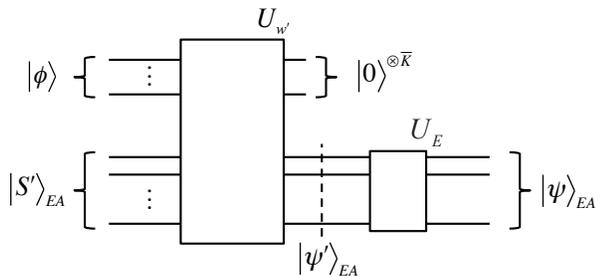}}
    \caption{Mapping circuit of EA-CWS code to map $K$ logical states to $n+c$ physical qubits in $K$-dimensional code space.}
  \end{center}
\label{fig:NAQE_mapper}
\end{figure}
%======================================================

We must now consider how to encode an arbitrary state $|\phi\rangle$ into the state $|\psi'\rangle_{EA}$ in the code space spanned by linear combinations of the states $|w_l'\rangle$.  While there are more elegant ways of doing this, for clarity we will present an encoding based on a generalized SWAP operation.  We suppose that we have a $K$-dimensional system in a state
\begin{equation}
|\phi\rangle = \sum_{l=0}^{K-1}\alpha_l|l\rangle,\ \ \ \sum_{l=0}^{K-1} |\alpha_l|^2 = 1 ,
\end{equation}
where $\{|l\rangle\}$ is a standard basis.  We prepare our $n+c$ qubits in the initial state (\ref{initial_state}).  We then define a unitary transformation $U_{w'}$ that swaps the state $|\phi\rangle$ into the codeword:
\begin{equation}
U_{w'}\left( |\phi\rangle \otimes | S'\rangle_{EA} \right) = 
|0\rangle \otimes \sum_{l=0}^{K-1} \alpha_l  |w_l'\rangle \equiv |0\rangle\otimes|\psi'\rangle_{EA} .
\end{equation}

This generalized SWAP operation puts the encoded state into the codespace, but this is not adequate to provide error correction.  We follow it by an encoding unitary $U_E$, drawn from the Clifford group, which maps the stabilizer generators of the initial state given in (\ref{initial_generators}) to those of a CWS code in standard form.  Recall that these stabilizer generators correspond to the vertices of a graph.  Each stabilizer generator has a single $X$ operator on one qubit and a $Z$ operator on the qubits corresponding to neighboring vertices of the graph.  Extending this idea to an entanglement-assisted code, we will get generators of the following form:
\begin{eqnarray}
g_i &=& X_i Z^{\mathbf{r}_i} | I,\phantom{1}\phantom{1} \textrm{for } i=1,\cdots,(n-c) \nonumber\\
g_i &=& X_i Z^{\mathbf{r}_i} | Z_j,\phantom{1}\textrm{for }i=(n-c)+1,\cdots,n \nonumber\\
h_i &=& Z_{i} | X_j,\phantom{1}\textrm{for }i=(n-c)+1,\cdots,n
\label{encoded_generators}
\end{eqnarray}
where $j=i-(n-c)$ and $\mathbf{r}_i$ is the $i$th row vector of the adjacency matrix $A$ of the graph ${G}$.  After the unitary encoding operation $U_E$, the base state $|S'\rangle_{EA}$ is mapped to a new state $|S\rangle_{EA}$:
\begin{equation}
U_E|S'\rangle_{EA} = |S\rangle_{EA} .
\end{equation}
Similarly, the other basis states are mapped to basis states of a new codespace.  Word operators that act on the state $|S\rangle_{EA}$ to produce the other basis states of the code space are given by
\begin{equation}
w_l = U_Ew_l'U_E^\dagger,
\end{equation}
and the orthogonal basis states spanning the code space are
\begin{equation}
|w_l\rangle_{EA} = U_E | w'_l\rangle = w_l|S\rangle_{EA}.
\end{equation}
Fig.~\ref{fig:NAQE_mapper} shows the mapping circuit of a EA-CWS codes which encodes an arbitrary state $K$-dimensional state $|\phi\rangle$ into a codeword of $(n+c)$ physical qubits.

In an entanglement-assisted quantum code, it is assumed that errors do not occur on Bob's qubits, since they do not pass through the channel.  For a CWS code in standard form, all Pauli errors can be represented by $Z$ operators alone.  Using Eq.~(\ref{eq:CWS_map}), an error $Z^{\mathbf{v}}X^{\mathbf{u}}$ acting on Alice's qubits can be represented by a binary vector $Cl_G(Z^{\mathbf{v}}X^{\mathbf{u}})$.

Let's consider an example to clarify this.  Suppose an error $E=IXI|II$ occurs on a CWS code with $n=3$ and $c=2$.  The base state has a stabilizer generated by
\begin{eqnarray}
g_1 &=& XZZ|II, \nonumber\\
g_2 &=& ZXZ|ZI, \nonumber\\
g_3 &=& ZZX|IZ \nonumber\\
h_1 &=& IZI|XI,  \\
h_2 &=& IIZ|IX \nonumber.
\end{eqnarray}
These stabilizers are based on a simple ring graph of size 3.  Then, $E$ can be represented by a binary vector
\begin{equation}
\label{eq:ECWS_ex}
Cl_{\mathcal{G}}(E=IXI|II) = 101|10 .
\end{equation}
As stated above, the physical errors do not affect Bob's qubits.  However, as this example in  Eq.~(\ref{eq:ECWS_ex}) shows, when we covert to an effective error containing only $Z$ operators, this can include operators on Bob's side.

Once the stabilizer generators of the base state have been found and a desired set of correctable errors has been enumerated, one can search for word operators to produce a code that corrects those errors.  We first use the technique described above to covert all the errors to effective errors including only $Z$ operators, and represent them by a set of binary strings.  We then search for a set of classical binary codewords that can correct this error set.  These binary codewords correspond to word operators of a CWS code in standard form, which will include only $Z$ operators.  However, in the entanglement-assisted case there is a complication:  the word operators may include $Z$ operators on Bob's side.  The encoding operation must be applied only on Alice's qubits.  Therefore the $Z$ operators on Bob's side must be removed.  This is done by applying some combination of stabilizer generators to each word operator to cancel the $Z$ operators on Bob's side.

To be more concrete:  let $w_z$ be a word operator corresponding to a binary vector from a classical codeword, and let this word operator consist only of $Z$ and $I$ operator.  If $w_z$ has a $Z$ operator on Bob's $i$th qubit, then we multiply the word operator by the stabilizer generator having a $Z$ operator acting on Bob's $i$th  qubit to $w_z$ to the $Z$ on Bob side.  We do this repeatedly until we have eliminated all operators (except the identity) on Bob's side.  By this process, word operators can be constructed that act nontrivially on Alice side only. It is important to remark that the resulting word operators do not consist only of $Z$ operators, in general, unlike the word operators of standard CWS codes.

%%%%%%%%%%%%%%%%%%%%%%%%%%%%%%%%%%%%%%%%%
\subsection{Examples}

We now give two examples of some new entanglement-assisted non-additive codes based on our construction.  In each case we briefly sketch the construction method starting from the graph and the correctable error set.

%%%%%%%%%%%%%%%%%%%%%%%%%%%%%%%%%%%%%%%%%
\subsubsection{((5,16,2;1)) CWS code with entanglement}

A ((5,16,2;1)) code can be constructed from a simple ring graph with $5$ vertices, using one ebit.  The initial base state is 
\begin{eqnarray}
|S'\rangle_{EA}=|0000\rangle|\Phi_+\rangle.
\end{eqnarray}
After a unitary encoding operation $U_E$, the stabilizer generators of the encoded base state are
\begin{eqnarray}
g_1 = XZIIZ|I ,\ \ && \ \ g_2 = ZXZII|I , \nonumber\\
g_3 = IZXZI|I ,\ \ && \ \ g_4 = IIZXZ|I , \nonumber\\
g_5 = ZIIZX|Z ,\ \ && \ \ h = IIIIZ|X,
\end{eqnarray}
We want to detect all single-qubit Pauli errors on Alice's qubits (of which there are 15), for  a minimum distance of 2.  After using the above stabilizer operators to find effective error operators containing only $Z$ operators, the 15 corresponding classical binary errors are
\begin{eqnarray}
& &10000|0\quad01001|0\quad11001|0 \nonumber\\
& &01000|0\quad10100|0\quad11100|0 \nonumber\\
& &00100|0\quad01010|0\quad01110|0 \\
& &00010|0\quad00101|0\quad00111|0 \nonumber\\
& &00001|0\quad10010|1\quad10011|1. \nonumber
\end{eqnarray}
We can find a classical binary code that corrects this set of errors.  Its codewords are:
\begin{eqnarray}
\label{eq:ex_1bw}
& &00000|0\phantom{1}
00011|0\phantom{1}
00101|1\phantom{1}
00110|1\nonumber\\
& &01001|1\phantom{1}
01010|1\phantom{1}
01100|0\phantom{1}
01111|0\nonumber\\
& &10001|0\phantom{1}
10010|0\phantom{1}
10100|1\phantom{1}
10111|1\nonumber\\
& &11000|1\phantom{1}
11011|1\phantom{1}
11101|0\phantom{1}
11110|0.
\end{eqnarray}
So the dimension of the code space will be 16.  We use these binary codewords from Eq.~(\ref{eq:ex_1bw}) to construct the word operators $w_l$.  Since these operators must be applied only on Alice's qubits, we use stabilizer generators to eliminate $Z$ operators on Bob's side.  This gives the following set of word operators:
\begin{eqnarray}
IIIII|I\phantom{1}IIIZZ|I\phantom{1} && ZIZZY|I\phantom{1}ZIZIX|I\nonumber\\
ZZIZY|I\phantom{1}ZZIIX|I\phantom{1} && IZZII|I\phantom{1}IZZZZ|I\nonumber\\
ZIIIZ|I\phantom{1}ZIIZI|I\phantom{1} && IIZZX|I\phantom{1}IIZIY|I\nonumber\\
IZIZX|I\phantom{1}IZIIY|I\phantom{1} && ZZZIZ|I\phantom{1}ZZZZI|I.\nonumber
\end{eqnarray}
The word operators $w'$ for the initial base state $|S'\rangle_{EA}$ (before applying $U_E$) are
\begin{eqnarray}
IIIII|I\phantom{1}IIIXX|I\phantom{1} && IIXIY|I\phantom{1}IIXXZ|I\nonumber\\
IXIIY|I\phantom{1}IXIXZ|I\phantom{1} && IXXII|I\phantom{1}IXXXX|I\nonumber\\
XIIIX|I\phantom{1}XIIXI|I\phantom{1} && XIXIZ|I\phantom{1}XIXXY|I\nonumber\\
XXIIZ|I\phantom{1}XXIXY|I\phantom{1} && XXXIX|I\phantom{1}XXXXI|I.\nonumber
\end{eqnarray}
The above set of word operators for this code is not unique---there are other sets of word operators for a ((5,16,2;1)) code.

%%%%%%%%%%%%%%%%%%%%%%%%%%%%%%%%%%%%%%%%%
\subsubsection{((7,4,5;4)) CWS code with entanglement}

Our first example only allows error detection, not correction.  Still using a ring graph, but going to a larger number of qubits and ebits, we can construct a ((7,4,5;4)) non-additive quantum code within our framework. This code has minimum distance 5, so it can correct up to two single-qubit errors. The initial base state of this code, including $4$ ebits, is 
\begin{equation}
|S'\rangle_{EA}=|000\rangle|\Phi_+\Phi_+\Phi_+\Phi_+\rangle.\nonumber
\end{equation}
After applying a unitary encoding operator $U_E$ on Alice's qubits, we get stabilizer generators in standard form:
\begin{eqnarray}
g_1 = XZIIIIZ|IIII ,\ \ && g_2 = ZXZIIII|IIII , \nonumber\\
g_3 = IZXZIII|IIII ,\ \ && g_4 = IIZXZII|ZIII , \nonumber\\
g_5 = IIIZXZI|IZII ,\ \ && g_6 = IIIIZXZ|IIZI , \nonumber\\
g_7 = ZIIIIZX|IIIZ ,\ \ && h_1 = IIIZIII|XIII , \\
h_2 = IIIIZIII|IXII ,\ \ && h_3 = IIIIIZI|IIXI , \nonumber\\
h_4 = IIIIIIZ|IIIX .\ \  &&  \nonumber
\end{eqnarray}
For brevity we omit the list of classical binary errors equivalent to all one- and two-qubit errors on Alice's side (there are 210 of them).  The associated classical code correcting these effective errors is
\begin{eqnarray}
\label{eq:ex_2bw}
& &0000000|0000\phantom{1}1011110|1110\nonumber\\
& &1100010|1111\phantom{1}0011101|0001.
\end{eqnarray}
The codespace is 4-dimensional.  From Eq.~(\ref{eq:ex_2bw}), we construct the word operators $w_l$ of this code:
\begin{eqnarray}
IIIIIII|IIII\phantom{1} && ZIIXYXZ|IIII\nonumber\\
IZZYXYY|IIII\phantom{1} && ZIZZZZY|IIII.\nonumber
\end{eqnarray}
The corresponding word operators $w_l'$ that act on the initial state $|S'\rangle_{EA}$ are
\begin{eqnarray}
IIIIIII|IIII\phantom{1} && XIXYYYI|IIII\nonumber\\
XXIZZYZ|IIII\phantom{1} && IIXXXIY|IIII.\nonumber
\end{eqnarray}
Once again, this choice of word operators is not unique; there are other sets of word operator for the ((7,4,5;4)) code.

\section{Conclusions}

Using shared ebits between the sender and the receiver, we have presented a scheme to construct non-additive entanglement-assisted quantum error-correcting codes. Our framework is based on the standard form of codeword stabilized (CWS) codes, which are specified by a graph (which gives the structure of the stabilizer generators for a base state) and a classical binary code.  This code is chosen to correct an induced binary error model, which is obtained by applying stabilizer generators to a set of Pauli errors to produce effective errors containing only $Z$ and $I$ operators.  The word operators of the CWS code are determined by the codewords of the binary code.  Because of the use of shared entanglement, the effective error model can include errors on Bob's qubits, even though physical errors only affect Alice's qubits, since they are the only ones that pass through the channel.  Because the corresponding word operators must act only on Alice's qubits, we showed that the operators on Bob's side can be eliminated by applying appropriate stabilizer generators.  This means that the encoding operation can be done solely on Alice's side.  Finally we gave two example codes based on a ring topology:  a ((5,16,2;1)) error-detecting code and a ((7,4,5,4)) error-correcting code.

\begin{acknowledgments}
The authors thank Keith Chugg, Ching-Yi Lai, and Mark Wilde for useful conversations.  TAB acknowledges support from NSF Grants No.~CCF-0448658 and No.~CCF-0830801.
\end{acknowledgments}

\bibliography{EA_CWS}

%merlin.mbs apsrev4-1.bst 2010-07-25 4.21a (PWD, AO, DPC) hacked
%Control: key (0)
%Control: author (72) initials jnrlst
%Control: editor formatted (1) identically to author
%Control: production of article title (-1) disabled
%Control: page (0) single
%Control: year (1) truncated
%Control: production of eprint (0) enabled
\begin{thebibliography}{8}%
\makeatletter
\providecommand \@ifxundefined [1]{%
 \@ifx{#1\undefined}
}%
\providecommand \@ifnum [1]{%
 \ifnum #1\expandafter \@firstoftwo
 \else \expandafter \@secondoftwo
 \fi
}%
\providecommand \@ifx [1]{%
 \ifx #1\expandafter \@firstoftwo
 \else \expandafter \@secondoftwo
 \fi
}%
\providecommand \natexlab [1]{#1}%
\providecommand \enquote  [1]{``#1''}%
\providecommand \bibnamefont  [1]{#1}%
\providecommand \bibfnamefont [1]{#1}%
\providecommand \citenamefont [1]{#1}%
\providecommand \href@noop [0]{\@secondoftwo}%
\providecommand \href [0]{\begingroup \@sanitize@url \@href}%
\providecommand \@href[1]{\@@startlink{#1}\@@href}%
\providecommand \@@href[1]{\endgroup#1\@@endlink}%
\providecommand \@sanitize@url [0]{\catcode `\\12\catcode `\$12\catcode
  `\&12\catcode `\#12\catcode `\^12\catcode `\_12\catcode `\%12\relax}%
\providecommand \@@startlink[1]{}%
\providecommand \@@endlink[0]{}%
\providecommand \url  [0]{\begingroup\@sanitize@url \@url }%
\providecommand \@url [1]{\endgroup\@href {#1}{\urlprefix }}%
\providecommand \urlprefix  [0]{URL }%
\providecommand \Eprint [0]{\href }%
\providecommand \doibase [0]{http://dx.doi.org/}%
\providecommand \selectlanguage [0]{\@gobble}%
\providecommand \bibinfo  [0]{\@secondoftwo}%
\providecommand \bibfield  [0]{\@secondoftwo}%
\providecommand \translation [1]{[#1]}%
\providecommand \BibitemOpen [0]{}%
\providecommand \bibitemStop [0]{}%
\providecommand \bibitemNoStop [0]{.\EOS\space}%
\providecommand \EOS [0]{\spacefactor3000\relax}%
\providecommand \BibitemShut  [1]{\csname bibitem#1\endcsname}%
\let\auto@bib@innerbib\@empty
%</preamble>
\bibitem [{\citenamefont {Calderbank}\ \emph {et~al.}(1998)\citenamefont
  {Calderbank}, \citenamefont {Rains}, \citenamefont {Shor},\ and\
  \citenamefont {Sloane}}]{CRSS98}%
  \BibitemOpen
  \bibfield  {author} {\bibinfo {author} {\bibfnamefont {A.~R.}\ \bibnamefont
  {Calderbank}}, \bibinfo {author} {\bibfnamefont {E.~M.}\ \bibnamefont
  {Rains}}, \bibinfo {author} {\bibfnamefont {P.}~\bibnamefont {Shor}}, \ and\
  \bibinfo {author} {\bibfnamefont {N.~J.}\ \bibnamefont {Sloane}},\
  }\href@noop {} {\bibfield  {journal} {\bibinfo  {journal} {IEEE T. Inform.
  Theory}\ }\textbf {\bibinfo {volume} {44}},\ \bibinfo {pages} {1369}
  (\bibinfo {year} {1998})}\BibitemShut {NoStop}%
\bibitem [{\citenamefont {Gottesman}(1997)}]{GottesmanThesis}%
  \BibitemOpen
  \bibfield  {author} {\bibinfo {author} {\bibfnamefont {D.}~\bibnamefont
  {Gottesman}},\ }\href@noop {} {\emph {\bibinfo {title} {Stabilizer Codes and
  Quantum Error Correction}}}\ (\bibinfo  {publisher} {Caltech Ph.D. Thesis},\
  \bibinfo {year} {1997})\BibitemShut {NoStop}%
\bibitem [{\citenamefont {Cross}\ \emph {et~al.}(2009)\citenamefont {Cross},
  \citenamefont {Smith}, \citenamefont {Smolin},\ and\ \citenamefont
  {Zeng}}]{Cross:2009jo}%
  \BibitemOpen
  \bibfield  {author} {\bibinfo {author} {\bibfnamefont {A.}~\bibnamefont
  {Cross}}, \bibinfo {author} {\bibfnamefont {G.}~\bibnamefont {Smith}},
  \bibinfo {author} {\bibfnamefont {J.~A.}\ \bibnamefont {Smolin}}, \ and\
  \bibinfo {author} {\bibfnamefont {B.}~\bibnamefont {Zeng}},\ }\href@noop {}
  {\bibfield  {journal} {\bibinfo  {journal} {IEEE T. Inform. Theory}\ }\textbf
  {\bibinfo {volume} {55}},\ \bibinfo {pages} {433} (\bibinfo {year}
  {2009})}\BibitemShut {NoStop}%
\bibitem [{\citenamefont {Van~den Nest}\ \emph {et~al.}(2004)\citenamefont
  {Van~den Nest}, \citenamefont {Dehaene},\ and\ \citenamefont
  {De~Moor}}]{VandenNest04}%
  \BibitemOpen
  \bibfield  {author} {\bibinfo {author} {\bibfnamefont {M.}~\bibnamefont
  {Van~den Nest}}, \bibinfo {author} {\bibfnamefont {J.}~\bibnamefont
  {Dehaene}}, \ and\ \bibinfo {author} {\bibfnamefont {B.}~\bibnamefont
  {De~Moor}},\ }\href@noop {} {\bibfield  {journal} {\bibinfo  {journal} {Phys.
  Rev. A}\ }\textbf {\bibinfo {volume} {69}},\ \bibinfo {pages} {022316}
  (\bibinfo {year} {2004})}\BibitemShut {NoStop}%
\bibitem [{\citenamefont {Bennett}\ \emph {et~al.}(1996)\citenamefont
  {Bennett}, \citenamefont {DiVincenzo}, \citenamefont {Smolin},\ and\
  \citenamefont {Wootters}}]{Bennett96b}%
  \BibitemOpen
  \bibfield  {author} {\bibinfo {author} {\bibfnamefont {C.~H.}\ \bibnamefont
  {Bennett}}, \bibinfo {author} {\bibfnamefont {D.~P.}\ \bibnamefont
  {DiVincenzo}}, \bibinfo {author} {\bibfnamefont {J.~A.}\ \bibnamefont
  {Smolin}}, \ and\ \bibinfo {author} {\bibfnamefont {W.~K.}\ \bibnamefont
  {Wootters}},\ }\href@noop {} {\bibfield  {journal} {\bibinfo  {journal}
  {Phys. Rev. A}\ }\textbf {\bibinfo {volume} {54}},\ \bibinfo {pages} {3824}
  (\bibinfo {year} {1996})}\BibitemShut {NoStop}%
\bibitem [{\citenamefont {Bowen}(2002)}]{Bowen02}%
  \BibitemOpen
  \bibfield  {author} {\bibinfo {author} {\bibfnamefont {G.}~\bibnamefont
  {Bowen}},\ }\href@noop {} {\bibfield  {journal} {\bibinfo  {journal} {Phys.
  Rev. A}\ }\textbf {\bibinfo {volume} {66}},\ \bibinfo {pages} {052313}
  (\bibinfo {year} {2002})}\BibitemShut {NoStop}%
\bibitem [{\citenamefont {Brun}\ \emph {et~al.}(2006)\citenamefont {Brun},
  \citenamefont {Devetak},\ and\ \citenamefont {Hsieh}}]{Brun20102006}%
  \BibitemOpen
  \bibfield  {author} {\bibinfo {author} {\bibfnamefont {T.~A.}\ \bibnamefont
  {Brun}}, \bibinfo {author} {\bibfnamefont {I.}~\bibnamefont {Devetak}}, \
  and\ \bibinfo {author} {\bibfnamefont {M.-H.}\ \bibnamefont {Hsieh}},\
  }\href@noop {} {\bibfield  {journal} {\bibinfo  {journal} {Science}\ }\textbf
  {\bibinfo {volume} {314}},\ \bibinfo {pages} {436} (\bibinfo {year}
  {2006})}\BibitemShut {NoStop}%
\bibitem [{\citenamefont {Schlingemann}\ and\ \citenamefont
  {Werner}(2001)}]{Schlingemann01}%
  \BibitemOpen
  \bibfield  {author} {\bibinfo {author} {\bibfnamefont {D.}~\bibnamefont
  {Schlingemann}}\ and\ \bibinfo {author} {\bibfnamefont {R.}~\bibnamefont
  {Werner}},\ }\href@noop {} {\bibfield  {journal} {\bibinfo  {journal} {Phys.
  Rev. A}\ }\textbf {\bibinfo {volume} {65}},\ \bibinfo {pages} {012308}
  (\bibinfo {year} {2001})}\BibitemShut {NoStop}%
\end{thebibliography}%

\end{document}